\documentclass[12pt]{article}
\usepackage{a4}\usepackage{amsmath}\usepackage{amssymb}
\newcommand{\be}{\begin{equation}}
\newcommand{\ee}{\end{equation}}
\newcommand{\bea}{\begin{eqnarray}}
\newcommand{\eea}{\end{eqnarray}}
\newcommand{\beq}{\begin{eqnarray}}
\newcommand{\eeq}{\end{eqnarray}}
\newcommand{\beao}{\begin{eqnarray*}}
\newcommand{\eeao}{\end{eqnarray*}}
\newcommand{\nn}{\nonumber}
\newcommand{\pa}{\partial}
\newcommand{\la}{\lambda}
\newcommand{\E}{{\cal E}}
\newcommand{\om}{\omega}
\newcommand{\Ref}[1]{(\ref{#1})}
\newcommand{\al}{{\alpha}}
\newcommand{\ep}{{\epsilon}}
\newcommand{\el}{electromagnetic }
\begin{document}
\begin{center}

{\Large Reconsidering the quantisation of electrodynamics
with\\[
8pt]
boundary conditions and some measurable consequences}\\[30pt]

{\sc M. Bordag}\footnote{e-mail:
Michael.Bordag@itp.uni-leipzig.de} \\[12pt]
{\small  University of Leipzig, Institute for Theoretical Physics\\
  Augustusplatz 10/11, 04109 Leipzig, Germany}
\end{center}

\vspace{2cm}

\begin{abstract}
We show that the commonly known conductor boundary conditions
$E_{||}=B_\perp=0$ can be realized in two ways which we call
'thick' and 'thin' conductor. The 'thick' conductor is the
commonly known approach and includes a Neumann condition on the
normal component $E_\perp$ of the electric field whereas for a
'thin' conductor $E_\perp$ remains without boundary condition.
Both types describe different physics already on the classical
level where a 'thin' conductor allows for an interaction between
the normal components of currents on both sides. On quantum level
different forces between a conductor and a single electron or a
neutral atom result. For instance, the Casimir-Polder force for a
'thin' conductor is by about $13\%$ smaller than for a 'thick'
one.
\end{abstract}
\section{Introduction}\label{Sec1}
The quantisation of the \el field in the presence of conducting
boundaries is on the base of a broad area of physical phenomena
including the Casimir effect, the retarded interaction of atomic
systems with a conducting wall, their radiation properties in a
cavity and more which are frequently called {\it Cavity QED}. Many
of these phenomena are by now measured with high precision as for
example the Casimir effect and play an increasing role in
applications like nano-technology.

The interaction with a conducting boundary can be described by the
well known boundary conditions
\be\label{bc} E_{||}=H_\perp=0\ee
which follow from the mobility of the electrons in the boundary
surface $S$. The  present paper is based on the observation that
the conditions \Ref{bc} leave room for different boundary
conditions on the normal component $\E_\perp$ of the electric
field and, as a consequence, for different physics.

In the commonly adopted approach it is always assumed that
$\E_\perp$ satisfies a Neumann boundary condition on $S$. In
Coulomb gauge this follows after imposing a Dirichlet condition on
the electrostatic potential and corresponding conditions on the
transverse modes. This procedure was probably first used in
\cite{CP} to calculate the force acting on a neutral atom in front
of a conducting wall (Casimir-Polder force) and in \cite{Cas}
deriving the Casimir effect.

Motivated by the calculation of radiative corrections to the
Casimir force and in  order to have Lorentz invariance broken only
by the boundary in \cite{BRW} a quantisation of electrodynamics
had been developed where the conditions \Ref{bc} are implemented
in a 'minimal' manner. The point is that for the quantisation the
use of the electromagnetic potential is indispensable and the
boundary conditions  on the components of the electromagnetic
potential do not follow uniquely from \Ref{bc}. In \cite{BRW} the
path integral formulation in covariant gauge using a restriction
of the functional integration  space by delta functions on the
boundary surface such that the potentials obey \Ref{bc}  had been
used. A photon propagator $^SD_{\mu\nu}(x,y)$ was derived
satisfying \Ref{bc} and being defined in the hole space, i.e., on
both sides of the conducting surface which was assumed to be
infinitely thin. This propagator allowed for an easy calculation
of the radiative correction in \cite{BRW} for parallel planes and
in \cite{BoLi} for a sphere.

Below we show that the difference between both approaches can be
traced back to the boundary conditions on the normal component of
the electric field $\E_\perp$. Whereas  $\E_\perp$ fulfils a
Neumann conditions in the standard approach it remains without any
condition in the approach of \cite{BRW}. Both approaches describe
different physics which can be best described as a 'thick'
conductor versus a 'thin' one. For the 'thick' conductor one needs
to assume that inside there are electrons which can move freely in
the perpendicular direction making  the electric field vanish.
Then from Gauss law on the surface  only a Neumann condition is
possible for consistency reasons. In opposite, in the 'thin'
conductor one needs to assume that there is no freedom for the
electrons to move in the perpendicular direction either due to the
thinness or due to an anisotropic conductivity so that there is no
need to impose a condition on the normal component of the electric
field.

In terms of the electromagnetic potential, for a 'thick' conductor
we have  boundary conditions on all  components of the potential,
separated spectral problems on both sides and absolutely no
interaction across the conductor. For a 'thin' conductor in the
approach of \cite{BRW} which was in \cite{115} rewritten in terms
of new polarisations we have two polarisations with boundary
conditions and two without. The latter do not feel the boundary
and are the same as in free space. Below we show that from these
polarisations a classical force follows between charges on
different sides of the surface in case both are moving in the
normal direction. This makes the two setups describe different
physics.

The question appears about further measurable quantities which are
sensitive to the difference between both approaches. In QED the
first candidate is the Casimir effect. However,  in \cite{BRW} and
in subsequent papers the Casimir effect turned out to be the same.
The next candidate is the Casimir-Polder force. Below we show that
it is for a 'thin' conductor by $13\%$ smaller than for a 'thick'
one.

We note that differences between 'thin' and 'thick' conductors had
been reported earlier in \cite{RW1} and \cite{RW2} for the
fluctuations of the Casimir force between two plates where the
difference was in a mode propagating parallel to the plates.

We should stress that no such choice appears if the boundary is
considered in the mathematical sense, i.e.  when there is no
meaning of the space beyond the boundary.  Then one has to impose
boundary conditions on {\em all} components of the electromagnetic
potential. As was noticed in the context of quantum cosmology just
restricting the fields which satisfy the Coulomb gauge condition
may lead to an incorrect result (see,
\cite{Vassilevich:1994cz,Esposito:1997wt}),
\cite{Vassilevich:1998iz}).  In \cite{Vassilevich:1998iz} general
criteria had been formulated for the compatibility of gauge and
boundary conditions.

In the next section we discuss in detail the two sets of boundary
conditions, the corresponding mode expansions and propagators. In
the third section we calculate the Casimir-Polder force for a
'thin' conductor.   Discussions and conclusions are given in the
fourth section. In the appendix we display some sum rules used in
the text. Throughout the paper we use units with $\hbar=c=1$.

\section{Quantisation with boundary conditions}\label{Sec2}
In this section we consider the  quantisation of electrodynamics
with conductor boundary conditions which was introduced in
\cite{BRW}. The starting point is the path integral
representation,
\be\label{PIform}Z(j)=\int DA_\mu \ \prod\limits_{{x\in S , \ \nu}
}\delta\left(n^\mu F^*_{\mu\nu}\right) \ e^{iS(A)+iA_\mu j^\mu}
\ee
with  delta functions on the surface $S$ restricting the
integration space in the functional integral to such potentials
$A_\mu$ that the corresponding field strengths satisfy \Ref{bc}.
Here the potentials $A_\mu(x)$ are defined in the whole space,
i.e., on both sides of the boundary surface $S$. The action $S(A)$
contains the gauge breaking term $1/(2\alpha)(\pa_\mu A^\mu)^2$.
The delta function in \Ref{PIform} can be represented as a path
integral, $\prod\limits_{x\in S}\delta\left(n^\mu
F^*_{\mu\nu}\right)=\int  Db \ \exp\left(i\int_S dx \  b^\nu(x)
n^\mu F^*_{\mu\nu}(x)\right)$, over auxiliary fields $b^\nu(x)$
living on the surface $S$ after what the integration over $A_\mu$
is Gaussian and results in a new photon propagator
${^S}D_{\mu\nu}(x,y)$, see Eq. \Ref{SD} below, whose boundary
dependent part is quite different from the standard approach.
Although this had not yet been worked out in detail, the procedure
should be equivalent to a quantisation with constraints where the
auxiliary fields $b^\nu(x)$ play the role of Lagrange multipliers.

This propagator although derived in the framework of quantum field
theory can be used for the classical problem to find the field of
a given source. The Maxwell equation $\pa^\mu F_{\mu\nu}=j_\nu$ is
solved by
\be\label{sln} A_\mu(x)=\int dy \ {^S}D_{\mu\nu}(x,y) j_\nu(y)\ee
and the  field strengths which follow from \Ref{sln} obviously
fulfil \Ref{bc} and do not depend on the gauge in which
${^S}D_{\mu\nu}(x,y)$ was calculated if current conservation
$\pa^\mu j_\mu=0$ holds.

In order to make the discussion more transparent we turn to the
completely explicit expressions which can be written down for a
single plane surface $S$.  We assume it to be located in the point
$x^3\equiv z=0$. We work in Lorentz gauge with gauge parameter
$\al=1$ and use the polarisations $E_\mu^s$ which were introduced
in \cite{115} and the mode expansion
\be\label{me1}A_\mu(x)= \int dk_{||} \int dk_3 \
\frac{1}{\sqrt{2k_0}} \ \frac{e^{ik_\al x^\al }}{{2\pi}} \
\sum_{s=0}^3 E_\mu^s \ f_s(z,k_3) \ a_{s,k} \  + c.c. \ .\ee
Here the index $\al$ takes values $\al=0,1,2$ and $c.c.$ indicates
that the complex conjugated expression must be added. The
$a_{s,k}$ are the free coefficients in the classical expression
which become creation and annihilation operators when doing
quantisation. The mode functions $f_s(z,k_3)$ are discussed below.
The frequency is $k_0=\sqrt{k^2_{||}+k^2_3}$ where $k_{||}\equiv
\sqrt{k^2_1+k^2_2}$ is the momentum parallel to the plane. In this
representation the polarisations read
\be\label{E03}
E_\mu^0=\frac{1}{\Gamma}\left(\begin{array}{c}k_0\\k_1\\k_2\\0\end{array}\right)
, \ E_\mu^3=\left(\begin{array}{c}0\\0\\0\\1\end{array}\right), \
 E_\mu^1=\frac{1}{k_{||}}
\left(\begin{array}{c}0\\k_2\\-k_1\\0\end{array}\right), \
E_\mu^2=\frac{1}{\Gamma k_{||}}
\left(\begin{array}{c}k_{||}^2\\k_0k_1\\k_0k_2\\0\end{array}\right)
\ee
with $\Gamma=\sqrt{k_\al k^\al}\equiv\sqrt{k^2_0-k^2_{||}}$. They
form a basis,
\be\label{basis} g_{\mu\nu} = E_\mu^s \ g_{st} \ E_\nu^t , \ee
where $g_{st}=diag(1,-1,-1,-1)$ is a metric tensor like
$g_{\mu\nu}$. The third polarisation, $E_\mu^3$, is the normal
vector to $S$ and the polarisations with $s=1,2$ are transversal,
$k^\mu E_{\mu}^{s=1,2}=0$. In order to apply the boundary
conditions \Ref{bc} to \Ref{me1} we remark that they can be
written in terms of the dual field strength and take the form
$n^\mu F^*_{\mu\nu }=2\ep_{\mu\nu\la\delta}n^\mu\pa^\la A^\delta$.
Inserting the mode expansion \Ref{me1} it is obvious that the
polarisations with $s=0,3$ drop out and conditions only on the
mode functions $f_{s=1,2}(z,k_3)$ follow. Because the boundary
conditions and the  polarisations do not contain derivatives with
respect to $z$ these are Dirichlet conditions, i.e.,
\be\label{dir1}f_{s=1,2}(z,k_3)=0 \quad \mbox{for} \quad  z=0 .\ee
Then the corresponding mode functions are
\be\label{mf1}f_{s=1,2}(z,k_3)=\frac{\sin(k_3 z)}{\sqrt{\pi /2}}.
\ee
For the polarisations with $s=0,3$ we do not have boundary
conditions, hence they take the same form as in free space,
\be\label{mf2}f_{s=0,3}(z,k_3)=\frac{e^{ik_3 z}}{\sqrt{2\pi}} .\ee
In the mode expansion \Ref{me1} the integration over $k_{||}$ goes
over the whole $\mathbb{R}^2$. For the polarisations $s=0,3$ the
integration over $k_3$ goes also over the whole axis, $k_3\in
(-\infty,\infty)$, but for $s=1,2$ only over the half axis, $k\in
[0,\infty)$. Because of the boundary conditions, for $s=1,2$ we
have two completely separated problems on both sides of the
surface and, strictly, speaking, we should introduce independent
constants according to $a_{s,k} \to \Theta(-z) a_{-,s,k}+\Theta(z)
a_{+,s,k}$. The mode expansion \Ref{me1} has the property to solve
the Maxwell equations in the gauge $\al=1$ and to fulfil the
boundary conditions \Ref{bc}.

Next we consider the corresponding electric and magnetic fields
strengths. By means of $\E=\frac{\pa}{\pa x^0} A_i-\frac{\pa}{\pa
x^i} A_0$ we obtain from \Ref{me1}
\bea\label{me1E} \E_i&=& \int dk_{||} \int dk_3 \
\frac{1}{\sqrt{2k_0}} \ \frac{e^{ik_\al x^\al }}{{2\pi}}
\left\{\left(\begin{array}{c}0\\0\\1\end{array}\right)_{\!\!\!i}
ik_0(-a_{0,k}+a_{3,k})\frac{e^{ik_3 z}}{\sqrt{2\pi}} \nn \right. \\
&& + \left. \left(\begin{array}{c}k_2\\-k_1\\0\end{array}\right)_{\!\!\!i}
\frac{ik_0}{k_{||}} \frac{\sin(k_3 z)}{\sqrt{\pi/2}}a_{1,k}+
\left(\begin{array}{c}k_1k_3\sin(k_3 z)\\k_2k_3\sin(k_3
z)\\ik_{||}^2\cos(k_3 z)
\end{array}\right)_{\!\!\!i}\frac{ia_{2,k}}{k_{||}\sqrt{2\pi}}\right\}
\eea
and from $B_i=\ep_{ijk}\frac{\pa}{\pa x^j}A_k$ for the magnetic
field
\bea\label{me1B} B _i&=& \int dk_{||} \int dk_3 \
\frac{1}{\sqrt{2k_0}} \ \frac{e^{ik_\al x^\al }}{{2\pi}} \left\{
\left(\begin{array}{c}k_2\\-k_1\\0\end{array}\right)_{\!\!\!i}
i (-a_{0,k}+a_{3,k})\frac{e^{ik_3 z}}{\sqrt{2\pi}} \nn  \right. \\
\nn \\
&& + \left. \left(\begin{array}{c}k_1k_3\cos(k_3
z)\\k_2k_3\cos(k_3 z)\\-ik_{||}^2\sin(k_3 z)
\end{array}\right)_{\!\!\!i}\frac{a_{1,k}}{k_{||}\sqrt{2\pi}}
+ \left(\begin{array}{c}k_2\\-k_1\\0\end{array}\right)_{\!\!\!i}
\frac{-k_0}{k_{||}} \frac{\cos(k_3 z)}{\sqrt{\pi/2}}a_{2,k}
\right\}\eea
follows. It is seen that the normal component of the electric
field has a contribution which does not obey a boundary condition
although  the boundary conditions \Ref{bc} are obviously
fulfilled. The same holds for the parallel components of the
magnetic field.

The connection with the commonly used formulation can be
established by applying a Neumann condition to the normal
component of the electric field,
\be\label{Neum} \frac{\pa}{\pa z} \ \E_3 =0 \quad \mbox{for} \quad
z=0 .\ee
For this to hold we must take
\be\label{mf3}f_{0}(z,k_3)=i\frac{\sin(k_3 z)}{\sqrt{\pi /2}},
\quad f_{3}(z,k_3)=\frac{\cos(k_3 z)}{\sqrt{\pi /2}} \ee
in the mode expansion \Ref{me1} instead of \Ref{mf2} together with
the restriction of the integration $k_3\in[0,\infty)$.  After
that, by means of completeness of the polarisations $E_\mu^s$,
\Ref{E03}, the mode expansion \Ref{me1} must be equivalent to the
standard one which can be written in the form
\bea\label{me3} &&A_\mu(x)= \int dk_{||} \int_0^\infty dk_3 \
\frac{1}{\sqrt{2k_0}} \ \frac{e^{ik_\al x^\al }}{{2\pi}}
\\ && \times \left\{
\left(\begin{array}{c}1\\0\\0\\0\end{array}\right)\frac{\sin(k_3
z)}{\sqrt{\pi /2}} \ a_{0,k} +\left(\begin{array}{c}0\\k_1\sin(k_3
z)\\k_2\sin(k_3 z)\\-ik_3\cos(k_3
z)\end{array}\right)\frac{a_{3,k}}{k\sqrt{\pi /2}} \nn \right. \\
 && +\left.
\left(\begin{array}{c}0\\k_2\\-k_1\\0\end{array}\right)\frac{\sin(k_3
z)}{k_{||}\sqrt{\pi /2}} \ a_{1,k}+
\left(\begin{array}{c}0\\k_1k_3\sin(k_3 z)\\k_2k_3\sin(k_3
z)\\ik_{||}^2\cos(k_3
z)\end{array}\right)\frac{a_{2,k}}{kk_{||}\sqrt{\pi /2}} \right\}
+ c.c. \nn \eea
with $k=\sqrt{k_{||}^2+k_3^2}$. The equivalence is established by
the relations
\be\label{relat}a_2-\tilde{a}_2=
i\frac{k_{||}}{k}(\tilde{a}_0-\tilde{a}_3)=
-\frac{k_{||}}{k_3}(a_0-a_3),\ee
where the $a_s$ belonging to the mode expansion \Ref{me1} are
marked by a tilde.  All quantities which are calculated either in
the standard approach with the mode expansion \Ref{me3} or with
the 'new' polarisations $E_\mu^s$, \Ref{E03}, and the additional
condition \Ref{Neum} on the normal component of the electric
field, must deliver the same physical results. As we will discuss
in the next section, they do indeed. However, if we relax that
condition on $\E_\perp$  they will  not. In this way we
established that the mode expansion \Ref{me1} with the mode
functions \Ref{mf1} and \Ref{mf3} (instead of \Ref{mf2}) is
equivalent to the commonly used one so that the whole difference
is in the boundary condition \Ref{Neum} on the normal component of
the electric field.

In the remaining part of this section we collect some formulas
which we need for the calculation of the Casimir-Polder force in
the next section. The photon propagator ${^S}D_{\mu\nu}(x,y)$
 was derived in \cite{BRW} in general form and explicitly for
two parallel planes.  Here we restrict ourselves to the easier
case of one plane.  The propagator takes the form
\be\label{SD}{^S}D_{\mu\nu}(x,y)=D_{\mu\nu}(x-y)+
\overline{D}_{\mu\nu}(x,y), \ee
where
\be\label{Dfree} D_{\mu\nu}(x-y)=\int\frac{d^4k}{(2\pi)^4} \
\frac{e^{ik_\mu(x^\mu-y^\mu)}}{-k_0^2+\vec{k}^2-i\ep}
\left(g_{\mu\nu}-(1-\al)\frac{k_\mu k_\nu}{k^2}\right) \ee
is the (causal if $\ep>0$) photon propagator in free space with a
gauge parameter $\al$ and $\overline{D}_{\mu\nu}(x,y)$ is the
boundary dependent part,
\be\label{Db}\overline{D}_{\mu\nu}(x,y)=\int\frac{d^3k_\alpha}{(2\pi)^3}
\ \frac{e^{ik_\al(x^\al-y^\al)+i\Gamma\left( \mid x^3\mid +\mid
y^3\mid\right)}}{-2i\Gamma} \sum_{s=1,2} E_\mu^s E_\nu^s, \ee
with $\al=0,1,2$ and the polarisation vectors $E_\mu^s$  given by
\Ref{E03}.

Representation \Ref{Db} can be derived in the following way.
Consider the scalar propagator in free space \Ref{Dfree} without
the Lorentz structure and integrate over $k_3$,
\be\label{Dfreesc} D(x-y)=\int\frac{d^4k}{(2\pi)^4} \
\frac{e^{ik_\mu(x^\mu-y^\mu)}}{-k_0^2+\vec{k}^2-i\ep}
=\int\frac{d^3k_\alpha}{(2\pi)^3} \
\frac{e^{ik_\al(x^\al-y^\al)+i\Gamma\mid x^3-y^3\mid}}{-2i\Gamma}.
\ee
Next observe that the boundary conditions \Ref{bc} in terms of the
new polarisations are two Dirichlet conditions. The corresponding
scalar propagator can be expressed in the form
\be\label{DDir1} D^{\rm Dirichlet}(x,y)
=\int\frac{d^3k_\alpha}{(2\pi)^3} \
\frac{e^{ik_\al(x^\al-y^\al)}}{-2i\Gamma}\left(e^{i\Gamma\mid
x^3-y^3\mid}-e^{i\Gamma\left(\mid x^3\mid+\mid
y^3\mid\right)}\right). \ee
It is obvious that it fulfils the boundary condition in $x^3=0$
or in $y^3=0$. For both arguments on one side    the second
contribution is that of a mirror charge. For these arguments on
different sides,   for example $x^3>0$ and $y^3<0$, it is zero.

Finally, $\overline{D}_{\mu\nu}(x,y)$, \Ref{Db}, follows from the
second part in $D^{\rm Dirichlet}(x,y)$, \Ref{DDir1}, as the
projection on the two polarisations which are affected by the
boundary conditions by inserting the corresponding polarisation
vectors.

Let us note also for later use that $D^{\rm Dirichlet}(x,y)$,
\Ref{DDir1}, can be obtained from the more conventional
representation
\be\label{DDir2} D^{\rm
Dirichlet}(x,y)=\int\frac{dk_\al}{(2\pi)^3}
\int\limits_0^\infty\frac{dk_3}{\pi/2}
\frac{e^{ik_\al(x^\al-y^\al)}
\sin(k_3x^3)\sin(k_3y^3)}{-k_0^2+\vec{k}^2-i\ep} \ee
by integrating over $k_3$. Another well known representation
emerges if in \Ref{DDir2} integrating over $k_0$,
\be\label{DDir3}  D^{\rm
Dirichlet}(x,y)=i\int\frac{d^2k_{||}}{(2\pi)^2}
\int\limits_0^\infty\frac{dk_3}{\pi/2} \ \frac{1}{{2k_0}} \
e^{-ik_\al(x_\al-y_\al)}\sin(k_3x^3)\sin(k_3y^3)  \ee
with $k_0=\sqrt{k_{||}^2+k_3^2}$ which appears in this way
written in terms of the mode functions \Ref{mf1}. Note that in
opposite to \Ref{DDir1} the latter two representations are valid
only if both arguments are on one side of the boundary surface. We
conclude by writing the propagator \Ref{SD} in the form of
\Ref{DDir3}. Taking $\al=1$ and assuming $x^3$ and $y^3$ to be on
one side of the surface it reads
\bea\label{SDold}^SD_{\mu\nu}(x,y)&=&\int\frac{dk_\al}{(2\pi)^3} \
e^{ik_\al(x^\al-y^\al)} \nn \\ && \times \left\{
\int\limits_{-\infty}^\infty \frac{dk_3}{2\pi} \
\frac{e^{ik_3(x^3-y^3)}}{-k_0^2+k_{||}^2+k_3^2-i\ep}
 \sum\limits_{s,t=0,3} E_\mu^s \ g_{st} \ E_\nu^t   \right. \nn  \\
&& \left. + \int\limits_{0}^\infty \frac{dk_3}{\pi/2} \
\frac{\sin(k_3x^3)\sin(k_3y^3)}{-k_0^2+k_{||}^2+k_3^2-i\ep}
\sum\limits_{s,t=1,2} E_\mu^s \ g_{st} \ E_\nu^t \right\} .\eea
This propagator consists of two parts. The first one involving the
polarisations $s=0,3$ is the same as in free space and it is
defined in the hole space and the second one   involving the
polarisations $s=1,2$ is defined for both arguments on one side of
the boundary. For arguments on different sides it must be dropped.

For the boundary conditions of a 'thick' conductor a
representation of the propagator in parallel to \Ref{SD} can be
written down too. It starts from the observation that  the
polarisation $s=0$  according to \Ref{mf3} goes with a Dirichlet
condition and using \Ref{DDir1} we can simply include the second
contribution into $\overline{D}_{\mu\nu}(x,y)$ \Ref{Db}. For the
other polarisation, $s=3$, we have from \Ref{mf3} a Neumann
condition. The corresponding scalar propagator is given by Eq.
\Ref{DDir1} with a plus sign between the two exponentials. In this
way we can include it into $\overline{D}_{\mu\nu}(x,y)$ \Ref{Db}
in the same way as before only with a reversed sign. Summarising
we obtain the photon propagator for the 'thick' boundary
conditions for $\al=1$ in the form of ${^S}D_{\mu\nu}(x,y)$,
\Ref{SD}, with a boundary dependent part which is now given by
\be\label{Dbth2}\overline{D}_{\mu\nu}(x,y)=\int\frac{d^3k_\alpha}{(2\pi)^3}
\ \frac{e^{ik_\al(x^\al-y^\al)+i\Gamma\left( \mid x^3\mid +\mid
y^3\mid\right)}}{-2i\Gamma} \left( -E_\mu^0 E_\nu^0-E_\mu^3
E_\nu^3+E_\mu^1 E_\nu^1+E_\mu^2 E_\nu^2\right). \ee
We wrote the sum over the polarisations explicitly to underline
the reversed sign for $s=3$ which is due to the Neumann condition
for this polarization. From the above derivation it follows that
${^S}D_{\mu\nu}(x,y)$ with this $\overline{D}_{\mu\nu}(x,y)$ is
just another representation of the standard photon propagator
which follows from the mode expansion \Ref{me3}.

In order to further discuss the physical meaning of the 'thin'
conductor we use the representation \Ref{SDold} of the propagator
to calculate the interaction between two currents, $j_1(x)$ and
$j_2(y)$, which may be on the same or on different sides of the
boundary. According to \Ref{sln} it is given by
\be\label{jj}\int j_\mu(z) \ ^SD_{\mu\nu}(z,z') \ j_\nu(z') \ dz \
dz' .\ee
For charges at different sides only the contributions from the
polarisations  $s=0,3$ are nonzero. First we consider the charges
to be static. Then the corresponding current has only a
($\mu=0$)-component and that is independent on time, i.e., on
$z_0$. Hence the integration over $z_0$ can be carried out
delivering  in \Ref{SDold} a delta function $\delta(k_0)$. This
gives as usual the   Greens function known from electrostatics.
But the only remaining ($\mu=0$)-component is that in $E_\mu^0$
which is proportional to $k_0$ so that the whole expression
\Ref{jj} vanishes. in this way a charge at rest does not interact
with a current on the other side. For reasons of Lorentz
invariance the same holds for a charge moving in parallel to the
plane. The situation changes only if both charges are moving in
perpendicular direction. In that case the interaction \Ref{jj}
between them across the surface is  nonzero. This must be
considered as a difference in physics between a 'thin' and a
'thick' conductor on the classical level. In the next section we
show that this is the case on the quantum level too.

The boundary conditions of  a 'thin' conductor may seem strange
from the point of view of the commonly known ones. We would like
to stress here that they can be used in a consistent way to derive
either a mode expansion or the corresponding photon propagator as
the above discussions have shown. They are based on the assumption
that the conductor is infinitely thin, which of course cannot be
realized in nature. The question whether it can be a good
approximation to some physical situation can be answered only by
investigation of interactions like \Ref{jj} above.

The Casimir effect as the force acting between conducting surfaces
deserves a special consideration with respect to the two sets of
boundary conditions because it is the same for both.  This was in
fact shown in \cite{BRW} where a 'thin' conductor was used and the
standard result for the Casimir force was re-obtained. Later in
\cite{BoLi} the same had been shown for a sphere. The point is
that the Casimir effect is quite insensitive because it makes use
only of the fact that two boundary dependent polarisations
($E_\mu^s$ with $s=1,2$) are present. In other words it counts the
degrees of freedom in opposite to the Casimir-Polder force where
the time-component plays a special role.

At this place we should remark that in the usual approach in
Coulomb gauge the Casimir force receives two contributions from
the two transversal photons which correspond to the polarisations
$s=1,2$ in \Ref{me3}. If using Lorentz gauge with the same
polarisations there are four contributions obeying boundary
conditions, namely in addition the time-like and the longitudinal
photon ($s=0,3$ in \Ref{me3}). The result should be a doubled
Casimir force. As it was shown in \cite{Amb} by demanding BRST
invariance the ghosts become boundary dependent too and give two
compensating contributions returning the Casimir force to its
previous value. Although the ghosts do not couple to physical
fields we have to stress that they may be important in other
physical situations (see \cite{Bordag:1997fh} and \cite{BoVass}
for a discussion regarding ghost fields in dielectrics).

The situation with the time-like and the longitudinal photons is
different when calculating the Casimir-Polder force. Here there are no
contributions from the ghosts because they do not couple to the
electron and it can be shown that if performing the calculation in
Lorentz gauge using the mode expansion \Ref{me3} the contributions
form the modes with $s=0$ and $s=3$ compensate each other.

Another moment to be mentioned is the relation to the force
between two dielectric bodies or between a charge and a dielectric
body. In that case there is automatically a condition on the
normal component of the electric field present ($\epsilon\E_z$
must be continuous) so that this case turns in the limit into a
'thick' conductor.
\section{Interaction of a charge with a conducting boun\-dary}\label{Sec3}

In this section we calculate the force acting on a single electron
or on an atom in front of a conducting boundary for a 'thin'
conductor, i.e., without imposing a boundary condition on the
normal component of the electric field. The case of a 'thick'
conductor, i.e., with the Neumann condition on the normal
component of the electric field, is matter of the standard
calculation using the mode expansion \Ref{me3} and can be found in
numerous papers. Besides the original one \cite{CP} we mention
\cite{Barton:1970hw} and successors, \cite{Milton}, \cite{Hinds}
and the book \cite{Milonni}. Here we show that the same result can
be obtained in terms of the polarisations $E_\mu^s$, Eq.
\Ref{E03}, with the mode functions \Ref{mf3} or, equivalently,
with the photon propagator \Ref{Dbth2}.

The calculation for a 'thin' conductor starts from  the non
relativistic Hamilton operator
\be\label{Hnr} H=\frac{\left(\vec{p}-e\vec{A}\right)^2}{2m}+e
A_0+V(x), \ee
for a particle in some potential $V(r)$ which is coupled to the
electromagnetic field $A_\mu$. Note that we have to keep $A_0$
because we are not working in Coulomb gauge. We divide $H$ into an
unperturbed part,
\be\label{H0} H_0=\frac{p^2}{2m}+V(x), \ee
and a perturbation. For the unperturbed problem,
\be\label{Hn} H_0 \psi_n(x)=E_n \psi_n(x),
 \ee
it is assumed that the $\psi_n(x)\equiv \mid n>$ form a basis and
that the expectation values $<n\mid p_i \mid n'>$ are finite. The
wave function is assumed to be centred in a point at distance $a$
from the surface.  To complete the setup one has to enlarge the
state space of the $\psi_n$'s by the photon states.

For a single electron in front of a conducting wall  the potential
$V(r)$ must be non electrostatic (a magnetic field in a Penning
trap for instance). It is needed in order to keep the electron
away from a direct contact with the surface.

For an atom in front of the conducting wall the potential $V(r)$
is that of the nucleus,
\be\label{Vnucl} V(x)=\frac{-e^2}{4\pi |x-x_N|}, \ee
which is  at the position $x_N=(0,0,a)$, i.e.,  at distance $a$
from the surface.

The perturbation is given by the operator
\be\label{Hpert}\Delta
H=-\frac{e}{m}\vec{A}\vec{p}-\frac{e}{2m}(\vec{p}\vec{A})+eA_0+
\frac{e^2}{2m}\vec{A}^2 +\Delta V(x), \ee
where in the second term in the r.h.s. $\vec{p}=-i\nabla$ acts on
$\vec{A}(x)$. We have to keep this term when not working in
Coulomb gauge.  For the setup with an atom
\be\label{DV} \Delta
V(x)=\frac{e^2}{4\pi\sqrt{x^2+y^2+(z+2a)^2}}\ee
is  the contribution of the  mirror charge of the nucleus. It
appears as an external field whereas the mirror charge of the
electron (taken alone or in the atom) comes in from the
interaction with the \el field in $\Delta H$. For a single charge
$\Delta V$ is not present.

The perturbations are calculated according to the well known rules
of quantum mechanics. From $\Delta V(x)$ we have a first order
perturbation simply by taking it in the unperturbed state,
\be\label{desN} \delta_{\rm es, nucl}=\frac{e^2}{4\pi} \
\left(\frac{1}{2a}+\frac{Q}{16a^3}+\dots \right) ,\ee
with $Q=<n\mid 2x_3^2-x_1^2-x_2^2\mid n>$. Also the next-to-last
term gives rise the a first order perturbation,
\be\label{d1} \delta_1= \frac{e^2}{2m} <n,0_\gamma\mid
\vec{A}^2\mid n,0_\gamma>, \ee
where $|0_\gamma>$ denotes the photon ground state, whereas the
other ones contribute in second order according to the well known
formula
\be\label{E2} \Delta E_n^{(2)}=\sum_{n'}\sum_{s,k} \frac{\mid <
n,0_\gamma\mid \Delta H^{(2)} \mid n',(s,k)_\gamma
>\mid^2}{E_n-(E_{n'}+\om)}, \ee
where $\Delta H^{(2)}$ is that part in $\Delta H$, \Ref{Hpert},
which is linear in $A_\mu$. Here the intermediate states $\mid
n',(s,k)_\gamma>$ include all one-photon states.

The \el potentials are given by  Eq. \Ref{me1} where the $a_{s,k}$
and their complex conjugate are now annihilation and creation
operators. The polarisations are given by \Ref{E03} and the mode
functions by \Ref{mf1} for $s=1,2$ and in dependence on the choice
of 'thick' or 'thin' boundary conditions by the mode functions
\Ref{mf3} resp. \Ref{mf2} for $s=0,3$.

In order to calculate the expectation value in the first order
contribution $\delta_1$, \Ref{d1}, we need the mode expansion
\Ref{me1} and keep only the boundary dependent contributions,
i.e., the polarisations $s=1,2$. The vacuum expectation value
takes the form
\be\label{d1b} <0_\gamma\mid A_\mu(x)A_\nu(y) \mid 0_\gamma> =
\int {d^2k_{||}} \int\limits_0^\infty {dk_3} \frac{1}{\sqrt{2k_0}}
 \
\sum_{s=1,2}  \ E_\mu^sE_\nu^s  f_s(x,k_3) f_s^*(y,k_3).\ee
Using \Ref{DDir3}, \Ref{DDir1} and \Ref{Db} it can be expressed in
terms of the boundary dependent part of the propagator,
\be\label{d1a}\delta_1=\frac{e^2}{2m} \ \frac{1}{i}
<n|\overline{D}_{ii}(x,x)|n>,\ee
where the boundary independent part was dropped.  Using the
explicit form, \Ref{Db}, and $|x^3|=x^3$ because we use $x^3>0$
only, we obtain
\be\label{d1c}
\delta_1=\frac{e^2}{2m}\frac{1}{i}\int\frac{d^3k_\al}{(2\pi)^3} \
\frac{-e^{2i\Gamma a}}{-2i\Gamma}
\left(1+\frac{k_0^2}{\Gamma^2}\right) <n\mid e^{2i\Gamma
(x^3-a)}\mid n >. \ee
Here the factor $k_0^2/\Gamma^2$ comes from the second
polarisation ($s=2$). After a Wick rotation, $k_0\to ik_4$,
$\Gamma\to i\gamma\equiv i\sqrt{k_4^2+k_{||}^2}$ and expanding the
exponential, $\exp(2i\Gamma (x^3-a))=1+\dots$ (the electron wave
function is centred in $x^3=a$),
\bea\label{d1d} \delta_1&=-&\frac{e^2}{2m}\int\frac{d^3_{\rm E}
k}{(2\pi)^3} \ \frac{e^{-2\gamma a}}{2\gamma}
\left(1+\frac{k_4^2}{\gamma^2}\right) <n\mid 1+\dots) \mid n >\nn \\
&=& -\frac{e^2}{4\pi} \ \frac{1+1/3}{8\pi
ma^2}+O\left(\frac{1}{a^4}\right) \eea
follows, where in $(1+1/3)$ the '1' comes from the first
polarisation ($s=1$) (it is the same as in the standard approach)
whereas the '1/3' comes from $s=2$ (it is different).

The second order perturbation \Ref{E2} is
\be\label{d2a}\delta_2=\sum_{n'}\int\frac{d^2k_{||}}{(2\pi)^2}
\int\limits_0^\infty\frac{dk_3}{\pi/2}\sum_{s=1,2} \
\frac{1}{2\om} \ \frac{\mid<n\mid G_s\mid
n'>\mid^2}{E_n-(E_{n'}+\om )},
 \ee
where the sum over $s$ and the integration over $k$ come from the
intermediate photon states and the mode expansion \Ref{me1}
together with  the usual commutator relation  have been used. From
the first three terms in the r.h.s. of \Ref{Hpert} the matrix
elements become
\be\label{phiG}  G_s= \frac{e^{ik_tx^t}}{2\pi} \ f_s(x,k_3)\left(
\frac{-e}{m} E^s_t p_t +e E_0^s \right),   \ee
where \Ref{me1} and the mode functions \Ref{mf1} have been used.
Here and in the following the index $t$ takes values $t=1,2$ and
summation over repeated $t$'s is assumed.

Now we wish to use the new propagator
$\overline{D}_{\mu\nu}(x,y)$, Eq.\Ref{Db}. For this reason we
rewrite \Ref{d2a} by introducing the integration over $k_0$,
\be\label{d2b} \delta_2=\sum_{n'}\int\frac{d^2k_{||}}{(2\pi)^2}
\int\limits_0^\infty\frac{dk_3}{\pi/2}\sum_s
\int\limits_{-\infty}^\infty \frac{dk_0}{2\pi i} \
\frac{1}{-k_0^2+\vec{k}^2-i\ep} \ \frac{\mid<n\mid G_s\mid
n'>\mid^2}{-k_0+E_n-E_{n'}(1-i\ep)},
 \ee
where the integration has to pass the poles according to $\ep>0$.
This is in parallel to the relation between formulas \Ref{DDir2}
and \Ref{DDir3}.  Note that representation \Ref{d2b} is in this
form valid only if no intermediate antiparticle states are present
as it is the case in our non relativistic approximation. A
completely relativistic description had been given in
\cite{Bordag92} and in \cite{Bordag:1986vz} from which \Ref{d2b}
can be obtained in principle (it is quite tedious to do that in
detail).

We proceed by carrying out the integration over $k_3$ in the same
way as \Ref{DDir1} follows from \Ref{DDir2}, where in the factors
$\sin(k_3x^3)\sin(k_3y^3)$, which are contained in \Ref{DDir2},
the variable $x^3$ belongs to one of the scalar products $<n\mid
\dots \mid n'>$ and $y^3$ to the other one. Taking only the
boundary dependent part we obtain
\be\label{d2c} \delta_2=\sum_{n'}\sum_s  \ \frac1i
\int\frac{d^3k_\al}{(2\pi)^3} \frac{-e^{2i\Gamma a}}{-2i\Gamma} \
\frac{\mid<n\mid \tilde{ G}_s\mid
n'>\mid^2}{-k_0+E_n-E_{n'}(1-i\ep)} \ee
with the notation
$$ \tilde{G}_s=e^{ik_tx^t+i\Gamma (x^3-a)} \left( \frac{-e}{m} E^s_t
p_t +e E_0^s \right).$$
 Now the matrix elements must be
calculated. Taking into account  simplifications  like that
following in the integration over $k_\al$ from the symmetry under
rotations in the $k_t$-plane we obtain
\bea\label{M2}   \mid<n\mid \tilde{ G}_1\mid n'>\mid^2&=&
\frac{e^2}{m^2}\ \frac12 \sum_t\mid <n\mid  p_t\mid n'>\mid^2  \nn
\\
 \mid<n\mid \tilde{ G}_2\mid n'>\mid^2&=& \frac{e^2}{m^2}\
 \frac{k_0^2}{\Gamma^2} \ \frac12 \sum_t\mid <n\mid  p_t\mid
 n'>\mid^2 \nn \\ &&
+e^2 \ \frac{k_{||}^2}{\Gamma^2}\mid <n\mid \varphi \mid n'
>\mid^2
\eea
with
\be\label{wfct} \varphi=e^{ik_tx^t+i\Gamma (x^3-a)}   .\ee
Several dropped cross terms can be shown not to contribute to the
results below and in the contributions proportional to $p_t$ we
have put $\varphi=1$.

We proceed with doing approximations. Let us first consider a
single electron in front of the conducting plane. Here one uses
the so-called 'no-recoil' approximation which means that
$E_n-E_{n'}$ in the denominator of \Ref{d2c}, i.e., in the
electron propagator, is small,
\be\label{smalla}\frac{1}{-k_0+E_n-E_{n'}}=
-\frac{1}{k_0}-\frac{E_n-E_{n'}}{k_0^2}-\frac{(E_n-E_{n'})^2}{k_0^3}+\dots
\, . \ee
This corresponds to an expansion in powers of the ratio of the
distance $a$ to the transition wave lengths corresponding to the
energy levels $E_n$. The second term in the r.h.s. can be shown
not to contribute in the following for symmetry reasons. The wave
function $\varphi$, Eq. \Ref{wfct}, will be expanded in powers of
the momentum,
\be\label{expwf} \varphi=1+ ik_tx^t+i\Gamma (x^3-a)+\dots \; ,\ee
which corresponds to an expansion in powers of the ratio of the
radius of the orbit of the electron motion in the potential $V(x)$
in  Eq. \Ref{Hnr} to $a$. Both ratios are assumed to be small. We
obtain
\bea\label{nor1}&&\delta_2^{\rm
no-recoil}=\int\frac{d^3k_\al}{(2\pi)^3} \frac{-e^{2i\Gamma
a}}{-2i\Gamma} \ \frac{-1}{k_0} \left\{ -e^2 \
\frac{k_{||}^2}{\Gamma^2} \right. \\ && \left. +\sum_{n'}\left[
\frac{e^2}{m^2} \left(1+\frac{k_0^2}{\Gamma^2}\right)
\frac12\sum_t|<n|p_t|n'>|^2 \right.\right. \nn \\ && \left.\left.
+e^2 \ \frac{k_{||}^2}{\Gamma^2}
\left(k_{||}^2\frac12\sum_t|<n|x^t|n'>|^2-
\Gamma^2|<n|x^3-a|n'>|^2\right)(E_n-E_{n'})^2 \right] \right\}.\nn
\eea
Now we use the sum rule \Ref{px} and
$\Gamma=\sqrt{k_0^2-k_{||}^2}$ and arrive at
\bea\label{nor2}\delta_2^{\rm
no-recoil}&=&\int\frac{d^3k_\al}{(2\pi)^3} \frac{-e^{2i\Gamma
a}}{-2i\Gamma} \ \frac{-1}{k_0} \left\{ e^2 \
\left(-1+\frac{k_0^2}{\Gamma^2}\right) \right. \\
&& \left. +\frac{e^2}{m^2} \left[
\left(-1+\frac{2k_0^2}{\Gamma^2}+\frac{\Gamma^2}{k_0^2}\right)
\frac12\sum_t<p_t^2> +\left(-1+\frac{\Gamma^2}{k_0^2}\right)
<p_3^2> \right] \right\},\nn \eea
where we used $\sum_{n'} |n'><n'|=1$. The integration over $k_0$
can be carried out in the following way. First we note that in
order to make the approximation \Ref{smalla} we have to pass the
pole in $k_0=0$ from below (see Eq. \Ref{d2b}) at some finite
distance. Then we are left with the $k_0$-integrations
\be\label{type}\int\limits_{-\infty}^\infty \ \frac{dk_0}{2\pi}
\frac{f(k_0^2)}{k_0-i\ep} =\frac{i}{2}f(0) \quad\mbox{and}\quad
\int\limits_{-\infty}^\infty \ \frac{dk_0}{2\pi}
\frac{f(k_0^2)}{k_0^3-i\ep} =\frac{i}{2}f'(0).\ee
Hence in Eq.\Ref{nor2} the terms with a factor $k_0^2$ in the
numerator vanish. The terms with a factor ${\Gamma^2}/{k_0^2}$ can
be shown to vanish after carrying out the integration over
$k_{||}$. In this way we arrive at
\be\label{d2nr} \delta_{2~\rm thin}^{\rm no-recoil}=
\frac{e^2}{4\pi} \
\frac{1}{4a}\left(-1-\frac{1}{2m^2}\sum_t<p_t^2>
-\frac{1}{m^2}<p_3^2> \right),\ee
where we added the subscript 'thin'. Here the first contribution
in the r.h.s. is the pure electrostatic contribution from the
mirror charge of the electron, the  other two are the corrections
in the case of 'thin' boundary conditions.

Now we consider an atom in front of the plane. In $\delta_2$, Eq.
\Ref{d2c}, we want to perform the Wick rotation and assume $\mid n
>$ to be the ground state. In rotating the integration path of
$k_0$ only the pole in  $k_0=0$ from $E_{n'}=E_n$ comes close to
the integration path\footnote{If $\mid n >$  is not the ground
state the poles in $k_0=E_n-E_{n'}<0$ are crossed by the path and
give additional contributions which we do not consider here.}. We
assume the ground state to be non-degenerated. Than its
contribution to the sum over $n'$ can be calculated by the well
known  formula
\be\label{Vp} \int\limits_{-\infty}^\infty \ \frac{dk_0}{2\pi}
\frac{f(k_0)}{k_0-i\ep} =\frac{i}{2}
f(0)+\mbox{Vp}\int\limits_{-\infty}^\infty \
\frac{dk_0}{2\pi}\frac{f(k_0)}{k_0}.\ee
In this way $\delta_2$ divides into two
parts,
\be\label{d2d} \delta_2=\delta_2^{\rm pole}+\delta_2^{\rm VP},\ee
where $\delta_2^{\rm pole}$ comes from $n'=n$ in \Ref{d2c}  and
taking the pole part whereas $\delta_2^{\rm VP}$ comes from the
VP-integral.

In $\delta_2^{\rm pole}$ in this way we get rid of the
$k_0$-integration and of  the sum over $n'$ and obtain with
\Ref{M2}
\be\label{d2diag1} \delta_2^{\rm pole} =\frac12
\int\frac{d^2k_{||}}{(2\pi)^2} \frac{e^{-2k_{||} a}}{2k_{||}}
\left(\frac{e^2}{m^2} \frac12\sum_t\mid < n \mid  p_t \mid
n>\mid^2-e^2\mid < n \mid \varphi  \mid n>\mid^2\right).
 \ee
In taking the limes of large $a$  we expand $\varphi$ according to
\bea\label{}&& \mid < n \mid \varphi  \mid n>\mid^2  \nn = \mid <
n \mid \exp(ik_tx^t-k_{||}(x^3-a)) \mid n>\mid^2  \nn \\&&= \mid <
n \mid 1+ ik_tx^t-k_{||}(x^3-a) +\frac12 k_{||}^2(x^3-a)^2-\frac12
(k_tx^t)^2+\dots \mid n>\mid^2  \nn
\\&&=1+\frac12 Q+\dots \eea
and obtain
\bea\label{d2diag2} \delta_2^{\rm
pole}&=&-\frac{e^2}{2m^2}\int\frac{d^2k_{||}}{(2\pi)^2}
\frac{e^{-2k_{||} a}}{2k_{||}}\left(1+k_{||}^2 Q+\dots\right)\nn \\
&=&\frac{-e^2}{4\pi}\left(\frac{1}{4a}+\frac{Q}{16a^3}+\dots\right).
\eea
This pole part which involves not only the (00)-component of the new
photon propagator is the contribution of $k_0=0$, so it is
instantaneous and picks up just the electrostatic contribution.

For the considered case of an atom in front of the conducting
plane we observe that the first contribution in the last line in
\Ref{d2diag2} can be interpreted as the electrostatic interaction
of the electron with the mirror charge of the nucleus. It does not
depend on the electron state nor on a mass. Hence an identical
contribution must be present if considering the same problem with
electron and nucleus interchanged. Because we have to collect all
distance dependent contributions to the energy we must take that
contribution into account on the same footing as the first one. As
a consequence we have to double the first contribution in the last
line of \Ref{d2diag2}\footnote{A similar contribution to $Q$
exists but it is much smaller due to the larger mass.} and now it
cancels just the electrostatic contribution of the mirror charge
of the nucleus, $\delta_{\rm es,nucl}$, Eq. \Ref{desN}, within the
given approximation up to $O\left(\frac{1}{a^5}\right)$.

Now we consider the contribution of the VP-integral. After the
Wick rotation with $k_0\to ik_4$, $\Gamma\to i\gamma\equiv
i\sqrt{k_4^2+k_{||}^2}$, we obtain
\be\label{d2nd1} \delta_2^{\rm VP}=-\mbox{VP}\int\frac{d^3_{\rm
E}k}{(2\pi)^3} \frac{e^{-2\gamma a}}{2\gamma}
\sum_{n'}\sum_{s=1,2}\frac{\mid <n\mid \hat{G}_s\mid n'
>\mid|^2}{-ik_4+E_n-E_{n'}} \ee
with
\bea\label{Gh} &&\sum_{s=1,2} \mid <n\mid \hat{G}_s\mid n'
>\mid|^2 =\frac{e^2}{m^2}\left(1+\frac{k_{4}^2}{\gamma^2}\right)
\frac12\sum_t\mid<n\mid  p_t\mid n'
>\mid^2  \\
&&~~~~~~~~~~~~~~ -e^2  \frac{k_{||}^2}{\gamma^2}\left[
k_{||}^2\frac12\sum_t\mid<n\mid x_t\mid n'>\mid^2+
\gamma^2\mid<n\mid x^3-a\mid n'>\mid^2 \right] \nn \eea
which follows from Eq. \Ref{M2}.
Now we consider large $a$ and expand
\be\label{largea}
\frac{1}{-ik_4+E_n-E_{n'}}=\frac{1}{E_n-E_{n'}}+\frac{ik_4}{(E_n-E_{n'})^2}-
\frac{k_4^2}{(E_n-E_{n'})^3}+\dots \ .   \ee
Odd powers of $k_4$ do not contribute for symmetry reasons.
Inserting \Ref{Gh} and \Ref{largea} into \Ref{d2nd1} we arrive at
\bea\label{d2nd2} &&\delta_2^{\rm VP}=\int\frac{d^3_{\rm
E}k}{(2\pi)^3} \frac{e^{-2\gamma a}}{2\gamma} \nn \\
&&\times \Bigg\{
-\frac{e^2}{m^2}\left(1+\frac{k_{4}^2}{\gamma^2}\right)
\sum_{n'}\frac12\sum_t\frac{\mid<n\mid p_t\mid
n'>\mid^2}{E_n-E_{n'}} \\ &&+
\frac{e^2}{m^2}k_4^2\left(1+\frac{k_{4}^2}{\gamma^2}\right)
\sum_{n'}\frac12\sum_t\frac{\mid<n\mid p_t\mid
n'>\mid^2}{(E_n-E_{n'})^3}\nn \\ &&+ e^2\frac{k_{||}^2}{\gamma^2}
\sum_{n'} \frac{k_{||}^2\frac12\sum_t\mid<n\mid x_t\mid n'>\mid^2+
\gamma^2\mid<n\mid (x^3-a)\mid n'>\mid^2 }{E_n-E_{n'}}\Bigg\}.\nn
\eea
Next we   apply  sum rules \Ref{px} and \Ref{TRK}
 and  Eq. \Ref{al} for the static
polarizabilities $\alpha_i$ to the sums over $n'$.  In the
integration over $k$ we use spherical coordinates with
$\ep=\cos\theta$ and obtain
\bea\label{d2nd3} \delta_2^{\rm VP} &=&\frac{1}{4\pi^2}
\int_0^\infty dk \ k\int_0^1d\ep \ e^{-2ka} \ \ \Bigg\{
\frac{1+\ep^2}{m} \\ && - k^2\left[
\ep^2(1+\ep^2)\frac{\al_1+\al_2}{4}+\left(1-\ep^2\right)^2
\frac{\al_1+\al_2}{4}+\left(1-\ep^2\right) \frac{\al_3}{2} \right]
\Bigg\},\nn\eea
where the individual contributions have been written in the same
order as they appear in \Ref{d2nd2}. Finally we obtain
\be\label{d2nd4} \delta_2^{\rm VP}=\frac{e^2}{4\pi} \
\frac{1}{6\pi ma^2}- \frac{1}{32\pi^2a^4}\left(\frac25 \
(\al_1+\al_2)+ \frac25 \ (\al_1+\al_2)  +\al_3\right)
+O\left(\frac{1}{a^6}\right).\ee
Here the first contribution cancels just $\delta_1$, Eq.
\Ref{d1d}, and the second one is what in the 'thin' boundary
condition  approach comes in place of the known Casimir-Polder
expression \Ref{} for the interaction energy of an atom with a
conducting wall,
\be\label{CPneu} \delta_{\rm CP ~thin}=-
\frac{1}{32\pi^2a^4}\left(\left[ \frac{\al_1+\al_2}{4} \right]+
\left[ \frac{11}{5} \frac{\al_1+\al_2}{4} +\al_3\right]\right).\ee
Here the first square bracket results from the polarisation $s=1$
(it is the same as from the TE-mode in the traditional approach)
and the second from $s=2$ which is different. For a spherically
symmetric state with all $\al_i$ are equal we obtain
\be\label{Vergleich} \delta_{\rm CP ~thin}=\frac{13}{15} \
\delta_{\rm CP ~thick}\equiv\frac{13}{15}\frac{-3\al}{32\pi^2a^4},\ee
i.e., a reduction by about $13\%$.

In the remaining part of this section we show that the result for
a 'thick' conductor which in the standard approach follows with
the mode expansion \Ref{me3} can be obtained using a mode
expansion in terms of the polarisations $E_\mu^s$, Eq. \Ref{E03},
with the mode functions $f_{s=0,3}(z,k_3)$ given by Eq. \Ref{mf3}
instead of \Ref{mf2} too. In this way for  'thick' conductor
boundary conditions both sets of polarisations, the traditional
one, Eq. \Ref{mf3}, and $E_\mu^s$, Eq. \Ref{E03} (or the use of
the photon propagator \Ref{Dbth2}) are equivalent.

The calculations follow exactly the same lines as before. In
addition to the matrix elements \Ref{M2} we have now
\bea\label{M0}   \mid<n\mid \tilde{ G}_0\mid n'>\mid^2&=& e^2 \
\frac{k_{0}^2}{\Gamma^2}\mid <n\mid \varphi \mid n'
>\mid^2+
\frac{e^2}{m^2}\ \frac{k_{||}^2}{\Gamma^2} \frac12 \sum_t\mid
<n\mid p_t\mid n'>\mid^2 \nn
\\
\mid<n\mid \tilde{ G}_3\mid n'>\mid^2&=&\frac{e^2}{m^2}\ <n\mid
p_3\mid n'>\mid^2
 \eea
and these enter all formulas like \Ref{d2a} to \Ref{d2c} with a
negative sign (cf. \Ref{Dbth2}). Then we have in addition to
\Ref{nor1} for a single electron
\bea\label{dd2nr1}&& \Delta\delta_2^{\rm no-recoil}=\frac1i
\int\frac{d^3k_\al}{(2\pi)^3} \frac{-e^{2i\Gamma a}}{-2i\Gamma} \
\frac{-1}{k_0} \left\{ -e^2 \ \frac{k_{0}^2}{\Gamma^2} \right. \nn \\
&& \left. +\sum_{n'}\left[ -e^2 \ \frac{k_{0}^2}{\Gamma^2}
\left(k_{||}^2\frac12\sum_t|<n|x^t|n'>|^2-\Gamma^2|
<n|x^3-a|n'>|^2\right) \right.\right. \nn \\ && \left.\left.  -
\frac{e^2}{m^2} \left(\frac{k_{||}^2}{\Gamma^2}
\frac12\sum_t|<n|p_t|n'>|^2+|<n|p_3|n'>|^2\right)\right] \right\}
. \eea
We apply the sum rules and rewrite
\be\label{dd2nr2}\Delta\delta_2^{\rm no-recoil}=\frac1i
\int\frac{d^3k_\al}{(2\pi)^3} \frac{-e^{2i\Gamma a}}{-2i\Gamma} \
\frac{-1}{k_0} \left\{ -e^2 \ \frac{k_{0}^2}{\Gamma^2}
-\frac{e^2}{m^2} 2\frac{k_{||}^2}{\Gamma^2}
\frac12\sum_t<p_t^2>\right\}
 .\ee
The contributions containing $<p_3^2>$ cancelled. Carrying out the
integrations we obtain finally
\be\label{ddnr}\Delta\delta_2^{\rm
no-recoil}=
\frac{e^2}{4\pi}\frac{1}{4a}\frac{1}{m^2}\sum_t<p_t^2>.\ee
Together with $\delta_{2~\rm thin}^{\rm no-recoil}$, Eq.
\Ref{d2nr} for the 'thin' conductor, these add up to the known
result for the 'thick' conductor,
\bea\label{nrsum}\delta_{2~\rm thin}^{\rm
no-recoil}+\Delta\delta_2^{\rm no-recoil}&=&\delta_{2~thick}^{\rm
no-recoil} \\ &=&
\frac{e^2}{4\pi}\frac{1}{4a}\left(-1+\frac{1}{2m^2}
\sum_t<p_t^2>-\frac{1}{m^2}<p_3^2>\right).\nn\eea

For an atom in front of the wall we make the approximation of
large $a$ and consider the additional contributions from the
polarisations $s=0$ and $s=3$. It can be shown that as before the
same leading contributions, i.e., that of order less than $1/a^4$,
cancel. The contributions to $\Delta\delta_2^{\rm VP}$, Eq.
\Ref{d2nd1},  which come in addition to \Ref{Gh}, read
\bea\label{M2a} && \sum_{s=0,3} \mid<n\mid \tilde{ G}_s\mid
n'>\mid^2\nn \\&&= e^2\frac{k_4^2}{\gamma^2}\left[k_{||}^2
 \frac12
\sum_t\mid <n\mid x_t\mid n'>\mid^2+
\gamma^2 \mid <n\mid x^3-a\mid n'>\mid^2\right]
\nn  \\ &&
 ~~~~~~ +\frac{e^2}{m^2}\left[ \frac{2k_{||}^2}{\gamma^2} \frac12
\sum_t\mid <n\mid p_t\mid n'>\mid^2-
\mid <n\mid p_3\mid n'>\mid^2\right].
\eea
Performing the same steps as in Eqs. \Ref{Gh} to \Ref{CPneu} we find
that the additional contribution is just
\be\label{}\Delta\delta_{\rm CP}=
\frac{-1}{32\pi^2a^4}\frac{4}{10}\frac{\al_1+\al_2}{2}
\ee
which together with $\delta_{\rm CP~thin}$, Eq. \Ref{CPneu}, gives
 the known result for the 'thick' conductor,
\be\label{}\delta_{\rm CP~thin}+\Delta\delta_{\rm CP}=
\delta_{\rm CP~thick}=-\frac{\al_1+\al_2+\al_3}{32\pi^2a^4}.
\ee
%

\section{Conclusions}\label{Sec4}
In the forging sections we have shown that the commonly known
conductor boundary conditions \Ref{bc} can be realized in two ways
which we denote as 'thick' and 'thin' conductors. For a 'thick'
conductor in addition to \Ref{bc} for the normal component of the
electric field on the surface the Neumann condition
\be\label{EZ} \frac{\pa}{\pa z} \E_z=0 \ee
holds. This is the commonly used approach. It can be formulated as
a mode expansion of the electromagnetic potential with the
standard polarisation vectors as shown in Eq. \Ref{me3}. The
boundary condition on $\E_z$ follows from considering the electric
field strength belonging to \Ref{me3},
\bea\label{Enor}   \E_i&=& \int dk_{||} \int dk_3 \
\frac{1}{\sqrt{2k_0}} \ \frac{e^{ik_\al x^\al }}{{2\pi}}
\left\{\left(
\begin{array}{c}k_1\sin(k_3z)\\k_2\sin(k_3z)\\-ik_3\cos(k_3z)
\end{array}\right)_{\!\!\!i}\frac{-i(a_{0,k}-a_{3,k})}{\sqrt{\pi/2}}
\nn \right. \\
&& + \left. \left(\begin{array}{c}k_2\\-k_1\\0\end{array}\right)_{\!\!\!i}
\frac{ik_0}{k_{||}} \frac{\sin(k_3 z)}{\sqrt{\pi/2}} \ a_{1,k}+
\left(\begin{array}{c}k_1k_3\sin(k_3 z)\\k_2k_3\sin(k_3
z)\\ik_{||}^2\cos(k_3 z)
\end{array}\right)_{\!\!\!i}\frac{i a_{2,k}}{k_{||}\sqrt{2\pi}}\right\}.
\eea
In terms of this mode expansion there is no choice not to take
$a_0$ in \Ref{me3} going with a $\sin(k_3z)$ because in Coulomb
gauge it is responsible for the electrostatic interaction with
must fulfil a Dirichlet condition. Now, if one wishes to have
$\nabla\E=0$ one needs for consistency reasons to take a
$\cos(k_3z)$ in the third component of the polarisation which goes
with $a_3$.

In this way, in the commonly used scheme it is natural to have the
condition \Ref{EZ} on the normal component of the electric field.
However, as demonstrated in the forgoing sections, there is a
choice not to do so. There exists a set of polarisation vectors,
$E_\mu^s$, Eq. \Ref{E03}, and a corresponding mode expansion, Eq.
\Ref{me1} which have the property to diagonalize the boundary
conditions in the sense that only the amplitudes belonging to two
polarisations ($E_\mu^s$ with $s=1$ and $s=2$) are affected by the
boundary conditions \Ref{bc} and the other two ($s=0$ and $s=3$)
are not. The modes with $s=0,3$ do not feel the boundary at all
and they are the same as without any boundary present. In this way
they are defined in the whole space and penetrate the boundary
freely. As a consequence we have to consider the whole space,
i.e., both sides of the boundary  which must be considered as
infinitely thin. Accordingly, the modes with $s=1,2$ are defined
separately on both sides (they do not penetrate). This setup we
have called a 'thin' conductor. The existence of a consistent
quantisation scheme for a 'thin' conductor had in fact been shown
in \cite{BRW} (even for a general, not flat surface) by the
construction of the corresponding photon propagator,
$^SD_{\mu\nu}(x,y)$ and in section 2 we considered the equivalent
mode expansion in detail.

In order to achieve a better understanding of the polarisations
$E_\mu^s$, Eq. \Ref{E03}, we considered in terms of these
polarisations the case of a 'thick' conductor by imposing the
Neumann boundary condition \Ref{EZ} in addition. Now also the
polarisations with $s=0,3$ become boundary dependent and the
corresponding mode functions are \Ref{mf3} in place of \Ref{mf2}.
In terms of the propagator we have the boundary dependent part now
given by E. \Ref{Dbth2} in place of \Ref{Db}. The equivalence to
the standard mode expansion \Ref{mf3} is established by Eq.
\Ref{relat}.

The difference between both types of conductors is a physical one
which can be seen already on classical level. Whereas a 'thick'
conductor does obviously not allow for any interaction between
currents, Eq. \Ref{jj}, on different sides of the boundary a
'thin' conductor does in case both currents have a component in
normal direction.

From this observation we conclude that a 'thick' conductor must
assume charges which can freely move inside in perpendicular (and
not only in parallel to the surface) direction. If there are no
such charges, either because of the thinness of the conductor or
due to an anisotropous conductivity, there is no need to impose
condition \Ref{EZ} on $\E_z$ and this is why we call it a 'thin'
conductor.

In section 3 we showed that   the forces acting on a single
electron or on an atom (Casimir-Polder force) in front of a
conducting wall are different for both types of conductors, for
instance the Casimir-Polder force is reduced by about $13\%$ for a
'thin' conductor whereas the Casimir force between two conducting
surfaces is the same.

The question whether a 'thin' conductor can be realized in
experiment is beyond the scope of this paper but one may think of
quasi two-dimensional systems like $C_{60}$ or other.

\section*{Acknowledgements}\label{Ack}
I thank D. Vassilevich and D. Robaschik for interesting
discussions and the University of Halmstad (Sweden), where part of
this work had been performed, for kind hospitality.

\section*{Appendix}\label{App}
Here we present the known sum rules which are used in the text:
\bea\label{px}\frac{i}{m}\ \frac{<n\mid p_i\mid n'> }{E_n-E_{n'}}
&=&<n\mid x_i\mid n'> ,\\
\label{TRK}\sum_{n'}\frac{\mid<n\mid p_t\mid n'>\mid^2
}{E_n-E_{n'}}&=&-\frac{m}{2} ,\\
\label{al} \frac{e^2}{m^2}\sum_{n'}\frac{\mid<n\mid p_i\mid
n'>\mid^2 }{\left(E_n-E_{n'}\right)^3}&=&-\frac{\al_i}{2} .\eea


\begin{thebibliography}{10}

\bibitem{CP}
H.B.G. Casimir and D.~Polder.
\newblock {The Influence of Retardation on the London-van der Waals Forces}.
\newblock {\em Phys.Rev.}, 73:360--372, 1948.

\bibitem{Cas}
H.B.G. Casimir.
\newblock {On the attraction between perfectly conducting plates}.
\newblock {\em Proc. Koninkl. Ned. Akad. Wetenshap}, 51:793, 1948.

\bibitem{BRW}
M.~Bordag, D.~Robaschik, and E.~Wieczorek.
\newblock {Quantum Field Theoretic Treatment of the Casimir Effect}.
\newblock {\em Ann. Phys.}, 165:192, 1985.

\bibitem{BoLi}
M.~Bordag and J.~Lindig.
\newblock {Radiative correction to the Casimir force on a sphere}.
\newblock {\em Phys. Rev.}, D58:045003, 1998.

\bibitem{115}
M.~Bordag.
\newblock {On the Canonical Quantization of QED with Boundary Conditions}.
\newblock 1984.
\newblock Preprint JINR-P2-84-115 (see KEK-scan).

\bibitem{RW1}
D.~Robaschik and E.~Wieczorek.
\newblock {Fluctuations of the Casimir pressure and the quantization of
  electrodynamics}.
\newblock {\em Ann. Phys.}, 236:43--68, 1994.

\bibitem{RW2}
D.~Robaschik and E.~Wieczorek.
\newblock {Fluctuations of the Casimir pressure at finite temperature}.
\newblock {\em Phys. Rev.}, D52:2341--2354, 1995.

\bibitem{Vassilevich:1994cz}
Dmitri~V. Vassilevich.
\newblock {QED on curved background and on manifolds with boundaries: Unitarity
  versus covariance}.
\newblock {\em Phys. Rev.}, D52:999, 1995.

\bibitem{Esposito:1997wt}
Giampiero Esposito, Alexander~Yu Kamenshchik, and Giuseppe
Pollifrone.
\newblock {\em Euclidean quantum gravity on manifolds with boundary}.
\newblock Kluwer, Dordrecht, 1997.

\bibitem{Vassilevich:1998iz}
Dmitri~V. Vassilevich.
\newblock {The Faddeev-Popov trick in the presence of boundaries}.
\newblock {\em Phys. Lett.}, B421:93--98, 1998.

\bibitem{Amb}
Jan Ambjorn and Richard~J. Hughes.
\newblock {Gauge fields, BRS symmetry and the Casimir effect}.
\newblock {\em Nucl. Phys.}, B217:336, 1983.

\bibitem{Bordag:1997fh}
M.~Bordag, K.~Kirsten, and D.~V. Vassilevich.
\newblock {Path integral quantization of electrodynamics in dielectric media}.
\newblock {\em J. Phys. A}, A31:2381, 1998.

\bibitem{BoVass}
M.~Bordag, K.~Kirsten, and D.~Vassilevich.
\newblock {On the ground state energy for a penetrable sphere and for a
  dielectric ball}.
\newblock {\em Phys. Rev.}, D59:085011, 1999.

\bibitem{Barton:1970hw}
G.~Barton.
\newblock Quantum electrodynamics of spinless particles between conducting
  plates.
\newblock {\em Proc. Roy. Soc. Lond.}, A320:251--275, 1970.

\bibitem{Milton}
Kimball~A. Milton, Jr. Lester L.~DeRaad, and Julian Schwinger.
\newblock {Casimir selfstress on a perfectly conducting spherical shell}.
\newblock {\em Ann. Phys.}, 115:388, 1978.

\bibitem{Hinds}
E.A. Hinds.
\newblock {Cavity Quantum Electrodynamics}.
\newblock {\em {Advances in Atomic, Molecular and Optical Physics}}, 28:237289,
  1991.

\bibitem{Milonni}
Peter~W. Milonni.
\newblock {\em The Quantum Vacuum: An introduction to quantum electrodynamics.}
\newblock Academic Press, 1993.

\bibitem{Bordag92}
M.~Bordag.
\newblock {Hydrogen levels between plates}.
\newblock In D.Yu. Grigoriev, V.A. Matveev, V.A. Rubakov, and P.G. Tinyakov,
  editors, {\em Seventh International Seminar on Quarks '92}, pages 80--94.
  World Scientific, 1993.
\newblock Preprint NTZ, University of Leipzig, 14/92.

\bibitem{Bordag:1986vz}
M.~Bordag.
\newblock {On the Apparatus Dependence of the Anomalous Magnetic Moment of the
  Electron}.
\newblock {\em Phys. Lett.}, B171:113, 1986.

\end{thebibliography}
\end{document}